\documentclass[%
 reprint,
 amsmath,amssymb,
 aps,
]{revtex4-2}

\usepackage{graphicx}
\usepackage{dcolumn}
\usepackage{bm}
\usepackage{braket}
\usepackage{color}
\usepackage{amsthm}

\DeclareMathOperator{\Tr}{Tr}

\usepackage{appendix}

\begin{document}

\preprint{APS/123-QED}

\title{Exploring the relationship between the faithfulness and entanglement of two qubits}

 \author{Gabriele Riccardi,$^{1}$ Daniel E. Jones,$^{2}$ Xiao-Dong Yu,$^{3}$ Otfried G{\"u}hne,$^{3}$ and Brian T. Kirby$^{2,4}$}
 \email{brian.t.kirby4.civ@mail.mil}
 \affiliation{
  $^1$ Department of Physical and Chemical Sciences  University of L’Aquila, 67100 L’Aquila, Italy\\
  $^{2}$ DEVCOM U.S. Army Research Laboratory, Adelphi, MD 20783, USA\\
  $^{3}$ Naturwissenschaftlich-Technische Fakultät, Universität Siegen, Walter-Flex-Straße 3, 57068 Siegen, Germany\\
  $^{4}$ Tulane University, New Orleans, LA 70118, USA
 }%

\date{\today}

\begin{abstract}
A conceptually simple and experimentally prevalent class of entanglement witnesses, known as fidelity witnesses, detect entanglement via a state's fidelity with a pure reference state.
While existence proofs guarantee that a suitable witness can be constructed for every entangled state, such assurances do not apply to fidelity witnesses.  Recent results have found that entangled states that cannot be detected by a fidelity witness, known as unfaithful states, are exceedingly common among bipartite states.  In this paper, we show that even among two-qubit states, the simplest of all entangled states, unfaithful states can be created through a suitable application of decoherence and filtering to a Bell state.
We also show that the faithfulness is not monotonic to entanglement, as measured by the concurrence.  
Finally, we experimentally verify our predictions using polarization-entangled photons and specifically demonstrate a situation where an unfaithful state is brought to faithfulness at the expense of further reducing the entanglement of the state.   
\end{abstract}

\maketitle


\section{Introduction}

Verifying the existence of entanglement in a quantum system is a problem of fundamental importance in quantum information science.  One potential approach is to perform quantum state tomography (QST) on the quantum system and use the obtained density matrix to determine separability.  However, due to the exponential resource requirements of QST and the fact that full state characterization is often not necessary for many use-cases, alternative methods of entanglement verification with lower overhead have been developed.

One prominent approach to verifying entanglement without full state characterization is to use an entanglement witness.  An entanglement witness is a Hermitian operator $W$ that has a positive expectation value for any separable state, but a negative expectation value for the entangled state of interest.  Therefore, given a target output state for a system, it is possible to construct an entanglement witness that will verify the entanglement of that state by measuring only a single expectation value.  

While various methods for constructing an entanglement witness exist, one of the most common is based on the fidelity of a state to a pure entangled state $\ket{\psi}$.  Fidelity $F_{\psi}(\rho)=\braket{\psi|\rho|\psi}$ is usually employed to measure the distance of a general state $\rho$ from the state $\ket{\psi}$. Starting from this, one can then build the related entanglement witness as \cite{bourennane2004experimental}
\begin{equation}
    W_{\psi}=\alpha \openone - \ket{\psi}\bra{\psi},
\end{equation}
where $\alpha$ is a suitably chosen real number. Measuring the observable $W$ leads to the quantity $\Tr{(\rho W)}= \alpha - F_\psi(\rho)$; if $F_\psi$ is above the threshold $\alpha$, the witness operator has detected the presence of entanglement. Clearly, one wants $\alpha$ to be as small as possible, so the witness can detect a larger amount of entangled states. Still, this method for entanglement detection is limited by the minimum value that $\alpha$ can have, and will thus have the drawback (common to every entanglement witness approach) of not being able to detect all entangled states. The term \emph{unfaithful} has been applied to the set of states that cannot be detected by any construction of a fidelity witness \cite{weilenmann2020entanglement,guhne2020geometry,zhan2020detecting}.  Recently, an analytical method for determining the faithfulness of any two-qubit state was 
provided \cite{guhne2020geometry}.  In order to be more explicit, we first express an arbitrary two qubit state $\rho_{AB}$ as
\begin{equation}
\begin{aligned}
    \rho_{AB}=&\frac{1}{4}\Big(\sigma_{0}\otimes\sigma_{0}+\vec{r}\cdot\vec{\sigma}\otimes\sigma_{0}+\sigma_{0}\otimes\vec{s}\cdot\vec{\sigma}\\
    +&\sum_{m,n=1}^{3}t_{nm}\sigma_{n}\otimes\sigma_{m}\Big),
    \label{eq:genstate}
\end{aligned}
\end{equation}
with the $\sigma_{i}$ representing the Pauli operators along with $\sigma_{0}=\openone$.
The faithfulness of $\rho_{AB}$ can then be determined using the operator
\begin{equation}
    X_{2}(\rho_{AB})=\rho_{AB}-\frac{1}{2}\left(\rho_{A}\otimes \sigma_{0}+\sigma_{0}\otimes\rho_{B}\right)+\frac{1}{2}\sigma_{0}\otimes \sigma_{0},
    \label{eq:x2def}
\end{equation}
where $\rho_{A,B}$ are the local density matrices.
A two-qubit state $\rho_{AB}$ is faithful if and only if the largest eigenvalue of $X_{2}(\rho_{AB})$ is greater than $1/2$ \cite{guhne2020geometry}.
States with completely mixed marginals, meaning they are Bell-diagonal in some basis, are only entangled when the largest eigenvalue of the state is greater than $1/2$ \cite{lang2010quantum}. 
Hence, for Bell-diagonal states in any basis, faithfulness and entanglement are one-to-one.
Therefore, only states with local correlations can potentially be entangled but unfaithful.

In this paper, we apply the analytical method of Ref.~\cite{guhne2020geometry} for determining the faithfulness of two-qubit states to Bell states undergoing both local filtration and decoherence.  
We then experimentally verify many of our results using polarization-entangled photons.  
As a central tool, we use a simple relationship between the fully entangled fraction (FEF) and the faithfulness of a two-qubit state, allowing for a physical interpretation of unfaithful two-qubit states as exactly those entangled states that are not useful for teleportation. 
Then, we show that a two-qubit entangled state, initially a Bell state, can be transformed into an unfaithful entangled state using only local filters and decoherence.  
We also demonstrate that the relative orientation of the filter and decohering elements is an essential factor in determining whether a state will become unfaithful.  
Further, we identify regimes where the faithfulness of a state and its entanglement, as quantified by the concurrence, behave in opposing manners, allowing for a trade-off between concurrence and faithfulness \cite{Albrecht2020maximal}. 
Specifically, we experimentally demonstrate a scenario where an unfaithfully entangled state of two qubits is brought to faithfulness at the expense of a further reduced concurrence.  
Our results demonstrate that a complex and often counter-intuitive relationship can exist between various metrics used to characterize quantum systems, even for the most basic entangled system of two qubits.

\section{Detecting and quantifying entanglement}
\label{sec:detecting}

In this section, we further develop the general concepts needed for the remainder of the paper.  In particular, we provide an overview of entanglement witnesses and the entanglement metric concurrence.  Further, we show that the faithfulness of a two-qubit state can be decided entirely from the FEF.  This allows us to offer a physical interpretation of unfaithful entanglement and hence identifies the set of states whose entanglement can be verified via a fidelity witness.

A bipartite quantum state is said to be separable if it can be written as
\begin{equation}
    \rho_{s}=\sum_{i=1}^{k}p_{i}\rho_{i}^{A}\otimes\rho_{i}^{B},
\end{equation}
for some probabilities $p_{i}$ and corresponding density matrices $\rho_{i}^{A}, \rho_{i}^{B}$.
If this is not the case, the state is entangled.
Even with knowledge of the full density matrix via a process like quantum state tomography, determining whether a given state is separable is NP-hard \cite{gharibian2010strong}.
Instead, many necessary conditions for this problem were proposed, such as the positive partial transpose (PPT) criterion, and the computable cross norm or realignment (CCNR) criterion, among others \cite{guhne2009entanglement}.

In principle, it is not necessary to obtain the entire density matrix of a state to decide separability.
As described in the introduction, an entanglement witness is a Hermitian operator $W$ constructed such that 
\begin{equation}
    \Tr{(W\rho_{s})}\ge0, \Tr{(W\rho_{e})}<0
\end{equation}
where $\rho_{s}$ is any separable state and $\rho_{e}$ is at least one entangled state \cite{guhne2009entanglement, terhal2002detecting, HORODECKI19961}.
It has been proven that for every entangled state $\rho_{e}$, it is possible to construct an entanglement witness \cite{HORODECKI19961}.
Despite the guaranteed existence of an entanglement witness and the myriad of construction methods, it is often challenging to realize a witness experimentally.
Due to their relative simplicity, fidelity based witnesses, as described in the introduction, are often the approach of choice in experimental scenarios.  

\begin{figure}[t]
\centering 
	\includegraphics[width=.9\columnwidth]{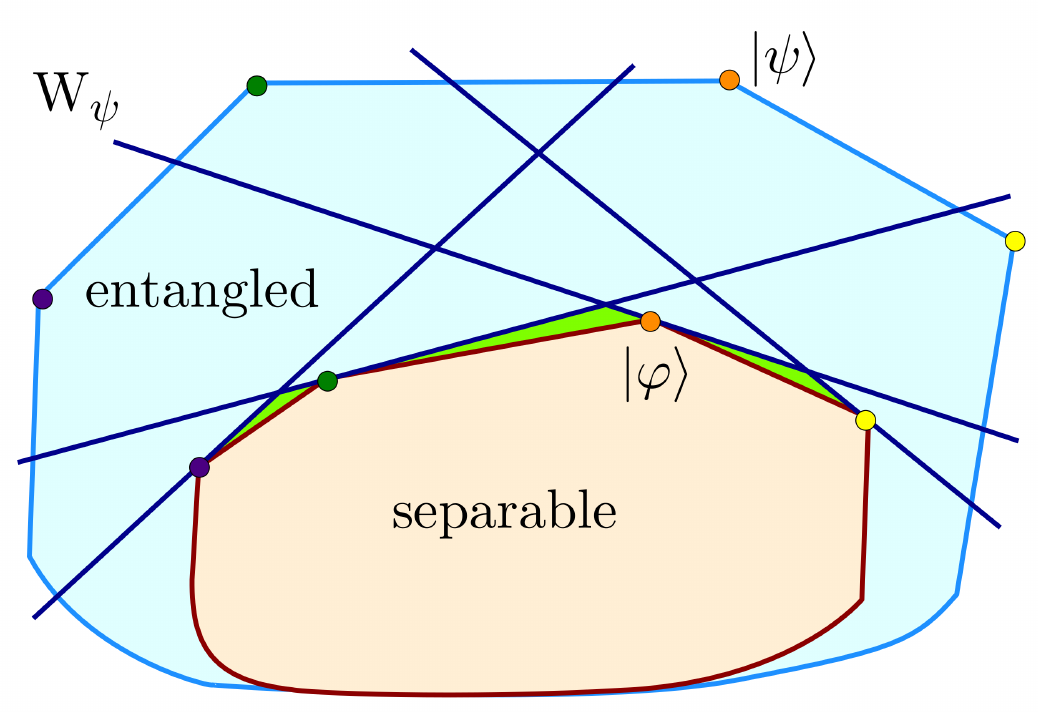}
	\caption{Schematic explanation of faithful and unfaithful entanglement. The set of all states is a convex set with the separable states as a convex subset; the extreme points of both sets correspond to pure states. 
	For constructing a fidelity-based witness, one starts with an entangled pure state $\ket{\psi}$ and computes the closest separable state $\ket{\varphi}$, where the distance is measured by the squared overlap. 
	Then, the witness $W_{\psi}$ is constructed by the condition
	that it should detect all states which have a smaller distance to $\ket{\psi}$ than the state $\ket{\varphi}$ has.
	The witness is depicted by the line in state space where $\Tr{(\rho W_\psi)} =0$, and the states above this line are detected as entangled. 
	The figure shows this procedure for four exemplary pure states. 
	States that can be detected by a fidelity-based witness are called faithful, but some entangled states (depicted in green) do not fall in this category and are hence unfaithful.}
	\label{Fig1-scheme}
\end{figure}

From a geometrical point of view, an entanglement witness is a hyperplane that is guaranteed to have all separable states on one side and at least one entangled state on the other \cite{guhne2009entanglement}; see also Fig.~\ref{Fig1-scheme}. 
With this picture in mind, we can see that a given entanglement witness will generally detect only a small subset of entangled states, namely those that fall on the same side of the hyperplane as the entangled state targeted in the construction of the witness. 
Fidelity-based witnesses also have a clear geometrical interpretation, as shown in Fig.~\ref{Fig1-scheme}. 
From this, it is also not surprising that not all entangled states can be detected by this specific class of witnesses, and such states are called unfaithful.

Beyond separability, the quality and type of entanglement present in a state determines how useful it is for various applications.  
For convenience of calculation in later sections, we adopt the concurrence, defined as $
\mathcal{C}(\rho)\equiv\max(0,\lambda_1-\lambda_2-\lambda_3-\lambda_4)$, to quantify entanglement.
Here, the $\lambda_i$ are in decreasing order and are the square roots of the eigenvalues of $\rho\tilde{\rho}$, where $\tilde{\rho}=(\sigma_{y}\otimes\sigma_{y})\rho^{*}(\sigma_{y}\otimes\sigma_{y})$ and $*$ denotes complex conjugation \cite{wootters2001entanglement}.
The concurrence itself does not have a simple physical interpretation.  However,  for two qubits, the entanglement of formation can be expressed as a monotonic function of the concurrence \cite{wootters2001entanglement}.

Finally, operationally motivated quantities, such as the FEF (sometimes referred to as the maximal singlet fraction), are also useful for characterizing quantum states.
The FEF is given by
\begin{equation}
    F(\rho)=\max_{\lvert \psi \rangle \in \text{ME}}
    \langle \psi \vert \rho \vert \psi \rangle, 
    \label{eq:FEFgen}
\end{equation}
where the maximization is over all maximally entangled (ME) states.  
The FEF is directly related to how useful a state is for teleportation.  
In the two-qubit case, only states having $F(\rho)>\frac{1}{2}$ can 
teleport a state with a higher fidelity than a classical channel \cite{horodecki1996teleportation,horodecki1999general}.
Despite the clear physical meaning of FEF, it is not an entanglement monotone and can show surprising behavior when compared with the concurrence \cite{Albrecht2020maximal}, a concept we will demonstrate experimentally below. 
Further, the FEF establishes upper and lower bounds on the concurrence and negativity \cite{grondalski2002fully,verstraete2002fidelity}.
We note that while FEF is in general difficult to evaluate as it requires a maximization, analytical results do exist for two qubits \cite{grondalski2002fully,bennett1996mixed,badziag2000local}, and estimates exist for higher-dimensional systems \cite{guhne2020geometry,yu2020quantuminspired}.
 
In fact, the maximum eigenvalue of the $X_2(\rho)$ operator from Eq.~(\ref{eq:x2def}) coincides with the FEF. 
This is central to our further discussion and, therefore,
is worth stressing: 

{\bf Observation.}
{\it
For two-qubit states, the maximum eigenvalue of the $X_2(\rho_{AB})$ 
operator coincides with the FEF of the state. This implies that an 
entangled state is unfaithful when $F(\rho_{AB})\le 1/2$.
}

The formal proof can be found in Refs.~\cite{guhne2020geometry, 
badziag2000local}, and for completeness, we also present a proof in the Appendix.

\section{Correlated faithfulness and concurrence}
\label{sec:corr}

In this section, we show how two-qubit unfaithful states can result from relatively simple applications of decoherence and filtering to Bell states. 
To begin, we consider a Bell state that has undergone a decoherence process that reduces the off-diagonal terms in the computational basis. This simple scenario is of practical interest since it can often be encountered in real-life situations. For example, it describes the case in which the initial Bell state is represented by a pair of polarization-entangled photons, and one or both photons propagate in a channel with first-order polarization-mode dispersion (PMD) \cite{antonelli2011sudden}.

The resulting state can then be represented as a Bell-diagonal state, which, up to local rotations, can be expressed in terms of the Pauli matrices as
\begin{equation}
\rho_0=\frac{1}{4}\left(\sigma_{0}\otimes\sigma_{0}+\sum_{j=1}^{3}t_{j}(\sigma_{j}\otimes\sigma_{j})\right).
\end{equation}
The values of the $t_j$ coefficients are constrained by the positivity conditions on the eigenvalues of $\rho_0$. 
Moreover, the Bell-diagonal state is separable if it obeys the stronger condition $\sum_{j}\vert t_{j}\vert \le 1$.

We now consider the application of a filter to this state. Such local filters 
can be written in terms of Pauli matrices as
\begin{equation}
\label{filter_op}
f = \mu\left(\sigma_{0}+\nu \hat{n}\cdot\vec{\sigma}\right),
\end{equation}
where $\mu$ and $\nu$ are real numbers and $\hat{n}$ is a unit vector that 
defines the direction of the filter in Stokes space. In the following, we 
neglect the parameter $\mu$, as we only consider the case where the state 
after filtration is normalized. Note that the filter $f$ is Hermitian, 
but this is no restriction since non-Hermitian filters can be represented 
by Hermitian filters followed by a local unitary transformation.

We refer to the local filter operator acting on qubit $A,B$ of the pair 
as $f_{A},f_{B}$. Application of this operator to qubit $A$ results in the 
unnormalized density matrix
\begin{equation}
\begin{aligned}
\rho'=&(f_{A}\otimes\sigma_{0})\rho_0(f_{A}\otimes\sigma_{0})^{\dagger}\\
=&f_{A}^{2}\otimes\sigma_{0}+\sum_{j=1}^{3}t_{j}(f_{A}\sigma_{j}f_{A}\otimes\sigma_{j}),\\
\end{aligned}
\end{equation}
where we have used the fact that $f$ is Hermitian and that the adjoint is distributive over the Kronecker product.
\newline
After algebra and normalization, this becomes
\begin{equation}
\label{eq_rho_f}
\begin{aligned}
\rho_{F}=&\rho'/\text{Tr}(\rho') = \frac{1}{4}\left(\sigma_{0}\otimes\sigma_{0}\right)+\frac{\nu}{2(1+\nu^{2})} \left[(\hat{n}\cdot\vec{\sigma})\otimes\sigma_{0}\right]\\
+&\frac{\nu}{2(1+\nu^{2})}\left[\sigma_{0}\otimes(\left(T\hat{n}\right)\cdot\vec{\sigma})\right]\\
+&\frac{1}{4(1+\nu^{2})}\sum_{j=1}^{3}t_{j}\left((1-\nu^{2})\sigma_{j}+2\nu^{2}n_j(\hat{n}\cdot\vec{\sigma})\right)\otimes\sigma_{j},\\
\end{aligned}
\end{equation}
where $T$ is the correlation matrix (diagonal matrix of $t_j$ values).
The concurrence of $\rho_{F}$ is independent of $\hat{n}$ (since this direction
can be adjusted by local unitaries) and depends only on the magnitude of 
$\nu$ \cite{kirby2019effect,verstraete2001local}, 
according to:
\begin{equation}
\label{eq_conc_filter}
    C(\rho_{F})=C(\rho_0)\frac{\vert \text{det}(f)\vert}{\text{Tr}\left[(f^{\dagger}f\otimes\sigma_{0})\rho\right]}.
\end{equation}

Following Ref.~\cite{guhne2020geometry}, in order to determine when this state becomes unfaithful, the eigenvalues of the operator $X_{2}(\rho_{F})$ must be determined. In 
this case, $X_{2}(\rho_{F})$ is given by
\begin{equation}
\begin{aligned}
X_{2}&(\rho_{F})=\frac{1}{4}\left(\sigma_{0}\otimes\sigma_{0}\right)\\
+&\frac{1}{4(1+\nu^{2})}\sum_{j=1}^{3}t_{j}\left((1-\nu^{2})\sigma_{j}+2\nu^{2}n_j(\hat{n}\cdot\vec{\sigma})\right)\otimes\sigma_{j}, \\
\end{aligned}
\end{equation}
which can always be diagonalized via local rotations and expressed as \cite{badziag2000local}
\begin{equation}
X_{2}(\rho_{F})= \frac{1}{4}\left(\sigma_{0}\otimes\sigma_{0}\right)+ \sum_{i=1}^3 t_i'(\sigma_i \otimes \sigma_i).
\end{equation}
The eigenvalues $\lambda_i$ are functions of the coefficients of its correlation matrix $T'$ \cite{lang2010quantum}, according to
\begin{equation}
\label{eq_eigenvalues}
    \begin{aligned}
    \lambda_{1}=&\frac{1}{4}\left(1+t_{1}'-t_{2}'+t_{3}'\right),\\
    \lambda_{2}=&\frac{1}{4}\left(1+t_{1}'+t_{2}'-t_{3}'\right),\\
    \lambda_{3}=&\frac{1}{4}\left(1-t_{1}'+t_{2}'+t_{3}'\right),\\
    \lambda_{4}=&\frac{1}{4}\left(1-t_{1}'-t_{2}'-t_{3}'\right).\\
    \end{aligned}
\end{equation}

If the maximum eigenvalue of $X_2(\rho_F)$ is no greater than $1/2$, then the state is unfaithful. Equivalently, in Sec.~\ref{sec:detecting}, we explained that we can consider the FEF since it coincides with the maximum eigenvalue of $X_2(\rho_F)$.
To study how the faithfulness of a Bell-diagonal state is affected by modal filtering, we consider several orientations of the filter in the following sections.

\subsection{Creating unfaithful entanglement with a single local filter}

\begin{figure}\centering 
	\includegraphics[width=.99\columnwidth]{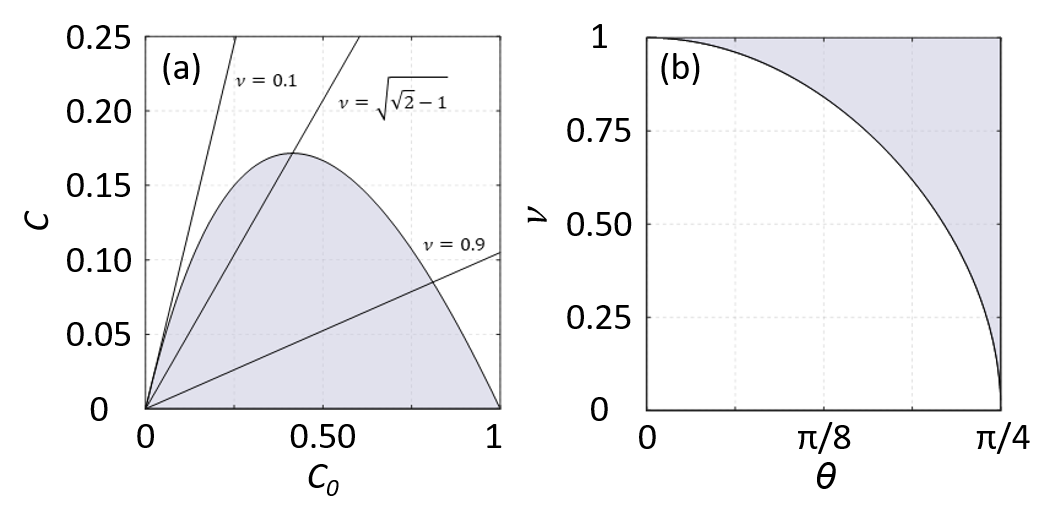}
	\caption{(a) Graphical representation of the inequality in Eq.~\eqref{condition}, where $C_0$ is the initial concurrence and $C$ is the concurrence obtained after application of a local filter on one of the two qubits. All combinations of $C_0$ and $C$ that lead to unfaithful entanglement are represented in the shaded area. Straight lines represent the linear relation between concurrence of the initial rank-2 Bell diagonal state and the concurrence of the final state after local filtering has been performed in Eq.~\eqref{c_vs_c0}. 
	(b) The same equality is expressed in terms of the amount of filtering $\nu$ needed for a state to become unfaithful when the initial Bell diagonal state is characterized by the coefficient $\theta$.}
	\label{Fig1}
\end{figure}

First, we focus on a simple but representative case, that is, the filter being aligned along the $x$-axis, such that $\hat{n}=\{1,0,0\}$.
In this situation, the coefficients $t_i'$ of the diagonal correlation matrix $T'$ become %
\begin{equation}
    \begin{aligned}
    t_{1}'=&\frac{2t_{1}\nu^{2}+t_{1}(1-\nu^{2})}{1+\nu^{2}}=t_{1},\\
    t_{2}'=&\frac{t_{2}(1-\nu^{2})}{1+\nu^{2}},\\
     t_{3}'=&\frac{t_{3}(1-\nu^{2})}{1+\nu^{2}},\\
    \end{aligned}
\end{equation}
which can be used to evaluate the FEF using Eq.~\eqref{eq:fef}.

Next, we simplify the description by assuming the initial Bell-diagonal state is rank two, so it can be described by the one-parameter density operator
\begin{equation}
\label{eq_r2bd}
    \rho_0=\sin^{2}(\theta)\vert \phi^{+}\rangle\langle\phi^{+}\vert +\cos^{2}(\theta)\vert\phi^{-}\rangle\langle\phi^{-}\vert,
\end{equation}
whose coefficients in the correlation matrix are $t_{1}=-\cos(2\theta)$, $t_{2}=\cos(2\theta)$, and $t_{3}=1$. 
The concurrence $C_{0}$ of rank-2 Bell diagonal states is given by $C_{0}=C(\rho)=|\cos(2\theta)|$.
Throughout the paper, we limit the discussion to the case in which $\theta$ is in the interval $0\le\theta<\pi/2$.

Using Eq.~\eqref{eq_conc_filter}, we see that the concurrence of this state after a local filter $f_{A}=(\sigma_{0}+\nu\sigma_{1})$ is applied is 
given by
\begin{equation}
\label{eq_concurrence}
    C=C(\rho_{F})=
    |\cos(2\theta)|\frac{1-\nu^{2}}{1+\nu^{2}},
\end{equation}
and we can use the relation
\begin{equation}
\label{c_vs_c0}
    \frac{C}{C_{0}}=\frac{1-\nu^{2}}{1+\nu^{2}}
\end{equation}
to express the FEF as
\begin{equation}
\label{eq:fef_one_filter}
    F(\rho_{F})=\frac{1}{4}\left(1+C_{0}+(C_{0}+1)\frac{C}{C_{0}}\right).
\end{equation}
Clearly, it can be inferred from Eq.~\eqref{eq:fef_one_filter} that the FEF is a monotonic function of the concurrence $C$ in this scenario.

We now want to investigate if this simple configuration can lead to the presence of unfaithful entanglement. 
As mentioned earlier, the condition for an entangled but unfaithful state is $C>0$ and $F(\rho_{F})\le1/2$.
Notice that if we start with a Bell state, meaning $C_{0}=1$, then $F(\rho_{F})=(1+C)/2$, and there are no situations where the state is unfaithful but also entangled, as expected.

From Eq.~\eqref{eq:fef_one_filter}, we see that the region where unfaithful entanglement occurs fulfills the condition
\begin{equation}
\label{condition}
    C\le\frac{C_{0}(1-C_{0})}{1+C_{0}},
\end{equation}
and is represented in Fig.~\ref{Fig1}(a) as the light blue area in the $C$ vs $C_0$ plane. Straight lines represent the linear relation between the concurrence of the initial rank-2 Bell diagonal state and the concurrence of the final state (after local filtering has been performed) in Eq. \eqref{c_vs_c0}. The figure shows that the highest concurrence for an unfaithful state that we can obtain by application of local filtering on a rank-2 Bell diagonal state is $C = 3-2\sqrt{2}$, which can only be obtained by applying a local filter of magnitude $\nu = \sqrt{\sqrt{2}-1}$ in any direction in the $\sigma_1$ - $\sigma_2$ plane to an initial state with concurrence $C_0=\sqrt{2}-1$.
Equation \eqref{condition} can also be rewritten to find an explicit relation between the amount of filtering required for the state to become unfaithful and the parameter $\theta$ characterizing the initial rank-2 Bell diagonal state to give
\begin{equation}
    \frac{1-\nu^2}{1+\nu^2} \le \frac{1-\cos{\theta}}{1+\cos{\theta}},
\end{equation}
which is represented in Fig.~\ref{Fig1}(b) (for $\theta$ in the range $0\le\theta<\pi/4$) by the shaded region. 
From this figure, it is clear that the larger the value of $\theta$ - i.e., the lower the initial concurrence of the rank-2 Bell diagonal state - the lower the amount of filtering needed to reach the region of unfaithful entanglement.

\subsection{Not all local filter orientations result in unfaithful entanglement}

Let's now consider a different filter orientation, which leads to very different results. 
If the filter is aligned such that $\hat{n}=\{0,0,1\}$, then the expression for $\rho_F$ becomes
\begin{equation}
\begin{aligned}
\rho_{F}=&\frac{1}{4}\left(\sigma_{0}\otimes\sigma_{0}\right)+\frac{\nu}{2(1+\nu^{2})} \left(\sigma_{3}\otimes\sigma_{0}\right)\\
&+\frac{t_{3}\nu}{2(1+\nu^{2})}\left(\sigma_{0}\otimes\sigma_{3}\right)+\frac{t_{3}\nu^{2}}{2(1+\nu^{2})}(\sigma_{3}\otimes\sigma_{3})\\
&+\frac{(1-\nu^{2})}{4(1+\nu^{2})}\sum_{j=1}^{3}t_{j}\left(\sigma_{j}\otimes\sigma_{j}\right).\\
\end{aligned}
\end{equation}

Again, if we assume that the initial state is a rank-2 Bell-diagonal state, we can analytically find the coefficients of the diagonal $T'$ matrix as
\begin{equation}
    \begin{aligned}
    t_{1}'=&\frac{t_{1}(1-\nu^{2})}{1+\nu^{2}}=-\frac{\cos{(2\theta)}(1-\nu^{2})}{1+\nu^{2}},\\
    t_{2}'=&\frac{t_{2}(1-\nu^{2})}{1+\nu^{2}}=\frac{\cos{(2\theta)}(1-\nu^{2})}{1+\nu^{2}}=-t_{1}',\\
    t_{3}'=&\frac{2t_{3}\nu^{2}+t_{3}(1-\nu^{2})}{1+\nu^{2}}=t_{3}=1.
    \end{aligned}
\end{equation}

The fully entangled fraction $F(\rho_{AB})$ in this case becomes (when $\theta$ is in the range $0 \le \theta < \pi/2$):
\begin{equation}
   F(\rho_{AB}) =\frac{1}{2}\left(1+\frac{C}{2}\right).\\
\end{equation}
Since this quantity is always greater than $1/2$, the state never becomes unfaithful.

This result shows how the faithfulness of a rank-2 Bell diagonal state on which a filter is applied is sensitive to the orientation of the filter itself. If the filter is aligned with the $z$-axis, one can never obtain an unfaithful state. This is in sharp contrast with the behavior of concurrence, which is orientation-independent.

\section{anti-correlated faithfulness and concurrence}
\label{sec:anticorr}

\begin{figure}\centering 
	\includegraphics[width=.99\columnwidth]{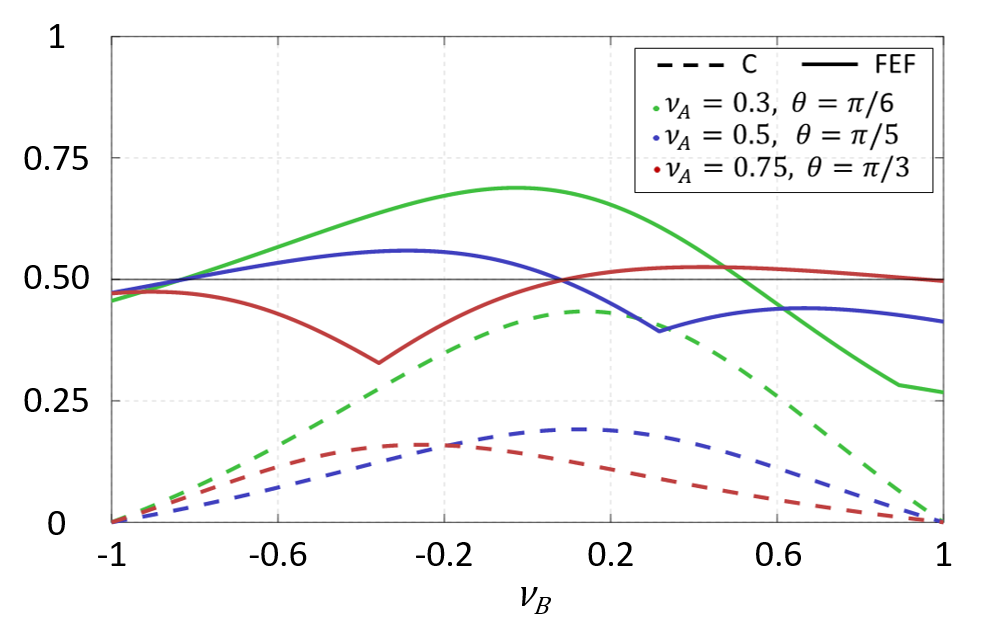}
	\caption{Theoretical curves for concurrence (dashed lines) and FEF (solid lines) as functions of the amount of filtering $\nu_B$ in channel B for three combinations of $\nu_A$ and $\theta$. A FEF above 0.5 (shown as a solid black line) means that the state is faithful. Filters in both channels are aligned along the \emph{x}-axis, which leads to the presence of regions where concurrence and FEF are anti-correlated.}
	\label{Fig2}
\end{figure}

 We now move to the description of a different scenario. Let's start again from the rank-2 Bell-diagonal state in Eq.~\eqref{eq_r2bd}. 
If we now apply a filter aligned in the $x$-direction to each qubit, so that the resulting operators are
\begin{equation}
f_{A} = \left(\sigma_{0}+\nu_{A} \sigma_{1}\right),\quad
f_{B} = \left(\sigma_{0}+\nu_{B} \sigma_{1}\right),
\end{equation}
the final state $\rho$ has a concurrence given by
\begin{equation}
   C(\rho_F)= \frac{\vert (\nu_{A}^2-1)( \nu_{B}^2-1)  \cos (2 \theta )\vert}{\left(\nu_{A}^2+1\right) \left(\nu_{B}^2+1\right)-4 \nu_{A} \nu_{B} \cos (2 \theta )},
\end{equation}
while, again limiting ourselves to the range $0 \le \theta < \pi/2$, the FEF becomes
\begin{equation}
\label{eq:fef_filterAB}
    F(\rho_F)=\frac{\max{\left[\sin ^2(\theta ) (\nu_{A} \nu_{B}+1)^2,\cos ^2(\theta ) (\nu_{A} \nu_{B}-1)^2\right]}}{\left(\nu_{A}^2+1\right) \left(\nu_{B}^2+1\right)-4 \nu_{A} \nu_{B} \cos (2 \theta )}.
\end{equation}

If we assume that the amount of filtering $\nu_A$ in channel A and $\theta$ are fixed, we can obtain by means of a simple derivative the value of $\nu_B^{\text{max}}$ for which the maximum FEF occurs. In fact, by setting $\frac{\text{d}F(\rho_F)}{\text{d}\nu_B}=0$ and solving for $\nu_B$, one has
\begin{equation}
\begin{aligned}
     F^{\text{max}}(\rho_F)=\max&\Big[\frac{(1+\nu_A^2)^2+4\nu_A^2\cos(2\theta)}{1+\nu_A^4-2\nu_A\cos{(4\theta)}}\sin^2{\theta},\\
     & \frac{(1+\nu_A^2)^2-4\nu_A^2\cos(2\theta)}{1+\nu_A^4-2\nu_A\cos{(4\theta)}}\cos^2{\theta}\Big].
\end{aligned}     
\end{equation}

When a filter is applied to channel B, the behavior of both concurrence and FEF strongly depends on the combination of the three parameters $\theta$, $\nu_A$ and $\nu_B$. 
In Fig.~\ref{Fig2}, we show theoretical plots for concurrence and FEF as functions of $\nu_B$ for the three different combinations of $\nu_A$ and $\theta$ specified in the legend. 
From the figure, one can clearly see that there are regions where concurrence and FEF behave oppositely. 
In particular, as the amount of filtering is increased in channel B, concurrence can decrease while, for a certain range of $\nu_B$, the FEF increases, even turning an unfaithful state into a faithful one.
This counter-intuitive behavior gives rise to a trade-off between concurrence and faithfulness and is experimentally verified in the next section.

\section{Experiment}

\begin{figure}\centering 
	\includegraphics[width=.99\columnwidth]{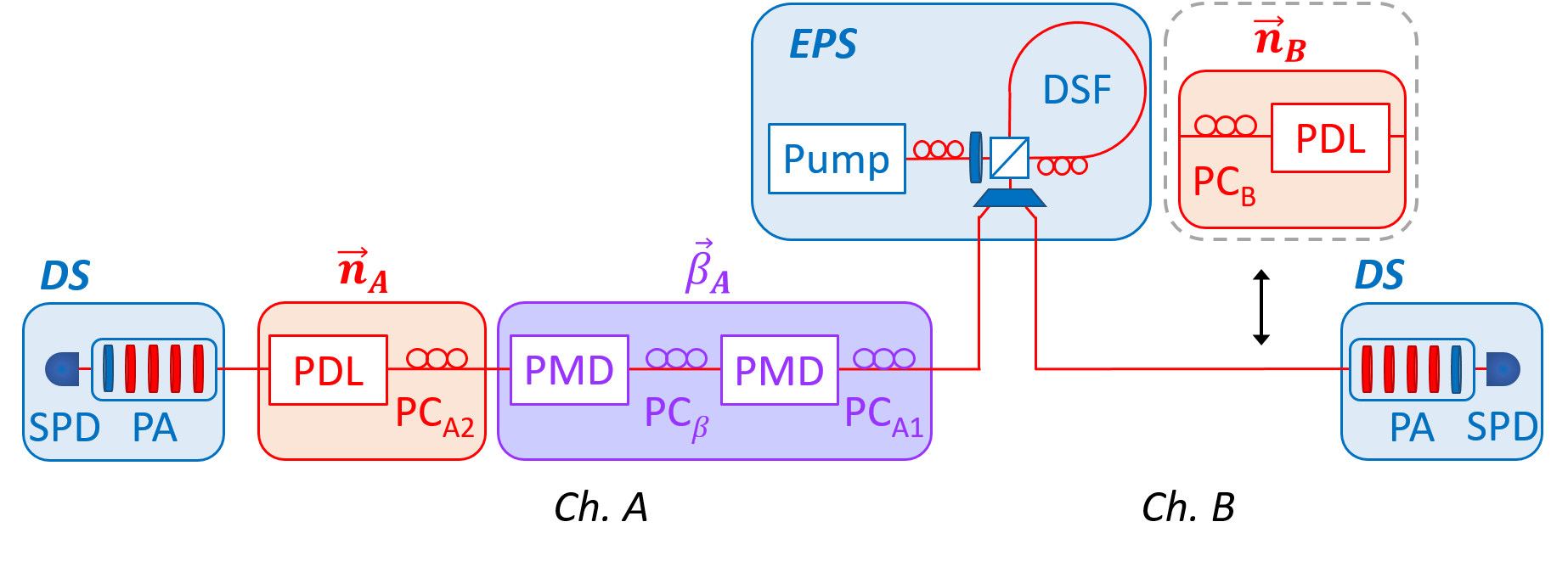}
	\caption{Schematic of the experimental apparatus. $\vec\beta_A$: birefringent element for the creation of a mixed Bell-diagonal state. $\vec n_A$: filtering element. $\vec n_B$: a second filtering element that can be introduced in the path of photon B. EPS: entangled photon source. DSF: dispersion-shifted fiber. PDL: PDL emulator. PMD: PMD emulator. PC: polarization controller. DS: detector station. PA: polarization analyzer consisting of several waveplates (red) and a polarizer (blue). SPD: single photon detector. }
	\label{experimental_setup}
\end{figure}
\begin{figure}[t]\centering 
	\includegraphics[width=.99\columnwidth]{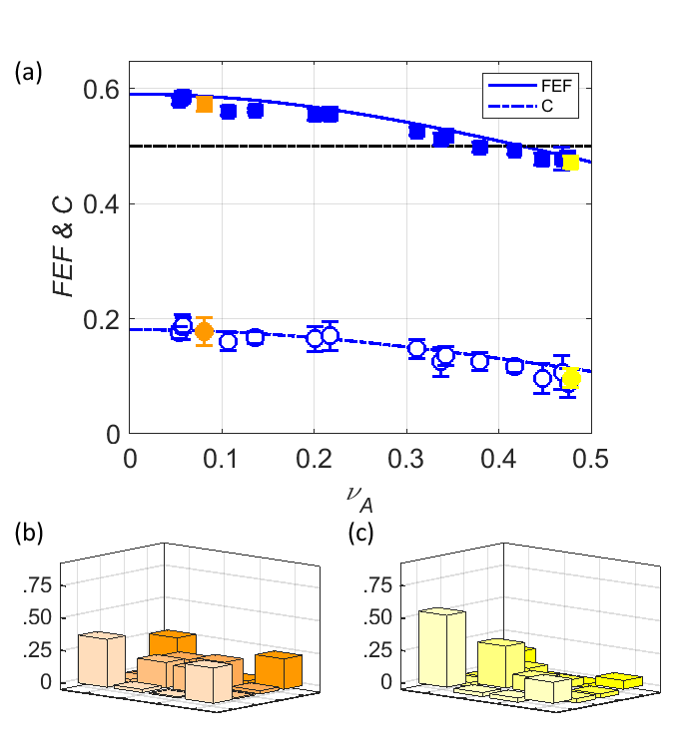}
	\caption{(a) Experimental data (markers) and theoretical curves for concurrence and FEF as the amount of filtering in channel A is increased. The filter is aligned along the direction $\hat n = \{1,0,0\}$ and is applied to photon A of a rank-2 Bell diagonal state. For $\nu_A \geq 0.426$, the state becomes unfaithful, but the nonzero concurrence shows that it is still entangled. Experimental density matrices corresponding to the orange and yellow markers are shown in (b) and (c), respectively, and are expressed in the basis $\vert HH \rangle, \vert HV \rangle, \vert VH \rangle, \vert VV \rangle$.}
	\label{expfig1}
\end{figure}
\begin{figure*}[!t]\centering 
	\includegraphics[width=1.99\columnwidth]{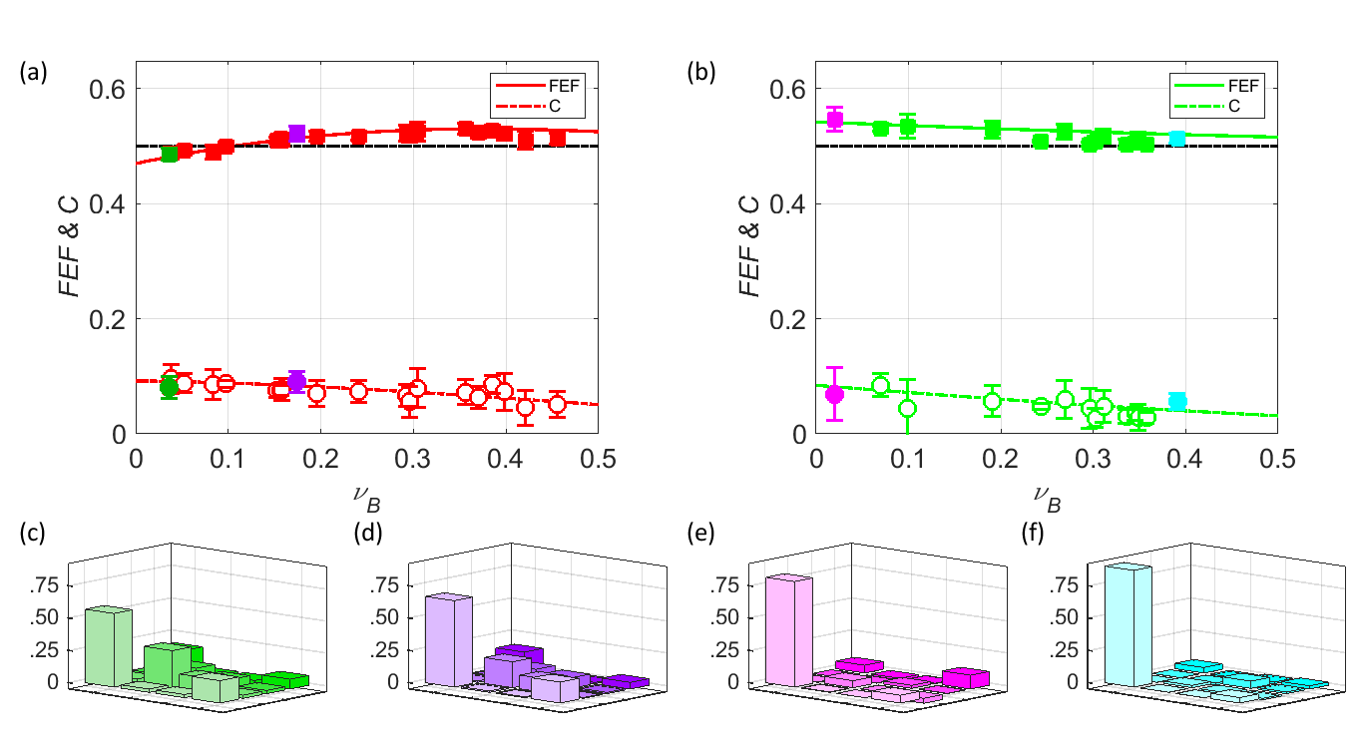}
	\caption{(a) - (b): Concurrence and FEF as functions of the amount ($\nu_B$) of filtering in channel B, while $\nu_A$ is kept fixed.
	The markers represent experimental results, while the curves are the theoretical predictions. 
	In (a), both filters are aligned along the direction $\hat n = \{1,0,0\}$ and $\nu_A = 0.465$ to demonstrate the anti-correlated behavior of concurrence and FEF. 
	In (b), both filters are aligned along the direction $\hat n = \{0,0,1\}$ and $\nu_A = 0.469$ to show a configuration where the state never becomes unfaithful. 
	The experimental density matrices in (c)-(f) refer to the markers identified by the same color. 
	The matrices in (c)-(d) are expressed in the basis $\vert HH \rangle, \vert HV \rangle, \vert VH \rangle, \vert VV \rangle$, and those in (e)-(f) are expressed in the basis $\vert RR \rangle, \vert RL \rangle, \vert LR \rangle, \vert LL \rangle$.}
	\label{expfig2}
\end{figure*}
A schematic diagram of our experiment is shown in Fig. \ref{experimental_setup}. 
Signal and idler photon pairs are generated via four-wave mixing \cite{fiorentino2002all} by pumping a dispersion shifted fiber (DSF) with a 50 MHz pulsed fiber laser which is filtered by a 100 GHz telecom add/drop filter centered at 1552.52 nm (ITU channel 31).
The pump laser is tunable, resulting in an average number of pairs per pump pulse in the range of $0.001 - 0.1$. 
The signal and idler photons are entangled in polarization, creating a $\ket{\phi^{+}}$ Bell state, by arranging the DSF in a Sagnac loop with a polarizing beam splitter (PBS) \cite{wang2009robust}.
The photons are then spectrally demultiplexed into 100 GHz-spaced ITU outputs after the Sagnac loop, and ITU channel 28 (1554.94nm) is sent to channel A, while ITU channel 34 (1550.12nm) is sent to channel B.
These filters result in photons with a temporal duration of about 15 ps. 
The detector stations (DS) each include one gated single photon detector (SPD) with a detection efficiency of $\eta \sim 20\%$ and a dark count probability of $ \sim 4 \times 10^{-5}$ per gate, as well as a polarization analyzer (PA) which allows for measurements at any angle on the Bloch sphere. 
FPGA-based controller software automatically controls the detectors and analyzers in order to perform full polarization state tomography by performing $36$ different measurements \footnote{Full state tomography of two qubits is achievable from the statistics of only $9$ measurement settings if two detectors are used per photon \cite{altepeter2005photonic}.  However, for convenience we use a single detector per photon and take $36$ total measurements, corresponding to all pairwise combinations of both eigenstates of each Pauli operator.}. 
Each of the 36 measurements is performed over 50 million detector gates, resulting in several thousands of detected coincidences per measurement depending on the experimental parameters.
The density matrix is then reconstructed using a maximum likelihood estimation algorithm \cite{altepeter2005photonic}, and the FEF and concurrence are calculated.
Due to the large amount of detected pairs per measurement, the resulting standard deviation of the FEF and concurrence (shown with the error bars in Figs. \ref{expfig1} and \ref{expfig2}) is rather small.

In order to demonstrate the case of correlated faithfulness and entanglement described in Sec.~\ref{sec:corr}, a decohering birefringent element $\vec\beta_A$ and a mode filter $\vec n_{A}$ are added to channel A. 
For polarization-entangled photons in optical fiber, polarization mode dispersion (PMD) acts as a form of decoherence, and polarization dependent loss (PDL) acts as a local filter \cite{antonelli2011sudden,brodsky2011loss,shtaif2011nonlocal,jones2018tuning}. 
As such, two PMD emulators with fixed differential group delay (DGD) and variable direction on the Bloch sphere are used to implement the decohering element $\vec\beta_A$ which transforms the nearly-perfect $\vert \phi^{+}\rangle$ state created by the EPS into a rank-2 Bell diagonal state \cite{kirby2019effect,jones2020exploring,riccardi2021simultaneous}.
The relative angle of the birefringence vectors applied by the two PMD emulators is adjusted using the polarization controller $PC_{\beta}$ to maximize the total DGD of the two elements, thus maximizing the decoherence of $\vec\beta_A$.
This results in an initial state (before filtering) with concurrence $C_{0} = .181$ which can reach the unfaithful regime with the application of an appropriate filter $\vec n_{A}$ since the initial concurrence is less than $C_{0} = \sqrt{2}-1$ as described following Eq. (14).
A PDL emulator is then used to implement the mode filter $\vec n_{A}$, whose magnitude and direction on the Bloch sphere are variable.

To achieve unfaithful entanglement, the filter $\vec n_{A}$ is aligned orthogonal to the decoherence vector $\vec\beta_A$ using the polarization controller $PC_{A2}$ \cite{jones2020experimental}, that is, the filter is aligned along the $x$-axis, such that $\hat{n}=\{1,0,0\}$ as in Sec. \ref{sec:corr}A.
The magnitude of the filter is increased from $\nu_{A} = 0.054-0.479$, and QST is performed for each value of $\nu_{A}$.
The FEF and concurrence of the resulting density matrices are plotted in Fig. \ref{expfig1}(a), and the density matrices corresponding to the orange and  yellow  markers are shown in Fig. \ref{expfig1}(b) and \ref{expfig1}(c), respectively \footnote{The matrices in Figs. \ref{expfig1}-\ref{expfig2} are expressed in the basis of the filter $\vec n_{A}$, and the $x$-axis in Stokes space is defined as $\vert H \rangle$. This basis allows for an intuitive understanding of how the filters $\vec n_{A}$ and $\vec n_{B}$ alter the density matrix; however, we note that the HV basis is rotated by $\pi/2$ (for both qubits) relative to the basis of the density matrix described by Eq. (16).}.
The behavior of the FEF and concurrence are clearly correlated as a function of $\nu_{A}$, and the state becomes unfaithful, while still entangled, for $\nu_{A} \geq 0.426$.

Next, additional measurements are performed to demonstrate the case where faithfulness and entanglement are anti-correlated, as described in Sec. \ref{sec:anticorr}.
The magnitude of the filter in channel A is fixed at $\nu_{A} = 0.465$, and an additional filter $\vec n_{B}$ is applied to channel B.
The filter in channel B is aligned in the same direction (along the $x$-axis) in Stokes space as the filter in channel A using $PC_{B}$, and its magnitude is increased from $\nu_{B} = 0.036-0.455$. 
Tomography is performed for each value of $\nu_{B}$, and the FEF and concurrence of the resulting density matrices are plotted in Fig. \ref{expfig2}(a).
The density matrices corresponding to the green and purple  markers are shown in Fig. \ref{expfig2}(c) and \ref{expfig2}(d), respectively.
The results clearly show that although entanglement always decreases as $\nu_{B}$ increases, the FEF increases over the range of $0 \le \nu_{B} \le 0.375$, and the state transitions from unfaithful to faithful at $\nu_{B} = 0.106$.
However, the state is always faithful, i.e. the FEF asymptotically approaches $0.5$ from above, when the filter in channel A is rotated such that it is collinear to the decoherence vector $\vec\beta_A$ (along the $z$-axis), and $\vec n_{B}$ is also rotated such that it remains in the same direction in Stokes space as $\vec n_{A}$.
For this scenario, the magnitude of the filter in channel A is fixed at $\nu_{A} = 0.469$, and the magnitude of the filter in channel B is increased from $\nu_{B} = 0.020-0.392$.
Tomography is once again performed for each value of $\nu_{B}$, and the results are shown in Figs. \ref{expfig2}(b),(e), and (f).

\section{Conclusion}

Two-qubit states are the most basic of all quantum systems that can be entangled. 
Despite this relative simplicity, two-qubit states underly many of the most important known quantum information technology applications, such as quantum key distribution and quantum teleportation. 
Hence, it is of great conceptual value to first explore new techniques and understand their properties before looking at higher-dimensions.  
In this paper, we showed that a surprisingly complicated relationship between entanglement and faithfulness emerges even for the common case of two-qubit states resulting from a Bell state that undergoes decoherence and filtering.
We also showed that entanglement and faithfulness could be traded, in some situations, with only a single local filter.  
These results provide an essential analogy for higher-dimensional systems where unfaithful entanglement is even more common.  
Further, even for two-qubit systems, our results have implications for quantum communication systems since the decoherence and filtering operations we consider in this paper are commonly found in optical networks.  

Several problems are suitable for further research. First, the counter-intuitive relation between faithfulness and entanglement should be studied further. 
The fact that the concurrence can decrease while the state becomes faithful may be interpreted to mean that sometimes highly entangled states are not necessarily easy to characterize in an experimental setting. 
This opens the question as to whether, in general, the complexity of the entanglement witness required to detect a state is a useful way to quantify entanglement.

Second, a natural question is how to expand the theory of faithfulness to the multi-particle setting.
For multi-particle systems, quantum state tomography becomes more demanding; hence entanglement characterization methods based on simple parameters (such as fidelities, the Fisher information, or spin-squeezing parameters) become more relevant. 
Still, it is not clear which types of quantum correlations can be characterized by these parameters and which cannot. 

\begin{acknowledgements}
BTK and GR would  like  to  acknowledge  valuable  discussions  with 
Hoi-Kwong Lo, Li Qian, and Yongtao Zhan. 

XDY and OG have been supported by the Deutsche Forschungsgemeinschaft (DFG, German Research 
Foundation, project numbers 447948357 and 440958198),
the Sino-German Center for Research Promotion,
and the ERC (Consolidator Grant 683107/TempoQ).
\end{acknowledgements}

\vspace{5mm}
\appendix
\section*{Appendix: Proof of the Observation}
To demonstrate the validity of the Observation, we start with the generic 
two-qubit state of Eq.~(\ref{eq:genstate}), where the correlation matrix, 
defined as the matrix made up of the coefficients
$t_{i,j}=\Tr{(\rho_{AB}\sigma_{i}\otimes\sigma_{j})}$ for nonzero $i,j$, 
can always be diagonalized via local rotations, resulting in
\begin{equation}
\begin{aligned}
    \rho_{AB}'=&\frac{1}{4}\Big(\sigma_{0}\otimes\sigma_{0}+\vec{r}'\cdot\vec{\sigma}\otimes\sigma_{0}+\sigma_{0}\otimes\vec{s}'\cdot\vec{\sigma}\\
    &+\sum_{m=1}^{3}t'_{m}\sigma_{m}\otimes\sigma_{m}\Big). \\
\end{aligned}
\end{equation}
Using this form, the FEF, as in Eq.~(\ref{eq:FEFgen}), can be evaluated 
as \cite{badziag2000local}
\begin{equation}
\label{eq:fef}
    F(\rho_{AB})= 
    \begin{cases}
        \frac{1}{4}(1+\sum_i |t'_i|) & \mbox{if} \det{(T')}\le 0\\
        \frac{1}{4}[1 + \max\limits_{i\ne j \ne k}{(|t'_i|+|t'_j|-|t'_k|)}] & \mbox{if} \det{(T')} > 0
    \end{cases}
\end{equation}
where $T'$ is a diagonal matrix with entries $t'_{i}$.
The operator $X_2(\rho_{AB})$, on the other hand, can be expressed as 
\begin{equation}
    X_2(\rho_{AB})=\frac{1}{4}\left(\sigma_{0}\otimes\sigma_{0}+\sum_{m,n=1}^{3}t_{nm}\sigma_{n}\otimes\sigma_{m}\right), 
\end{equation}
whose correlation matrix can again be diagonalized in the same way. 
One can thus explicitly determine the relationship between the eigenvalues of $X_2(\rho_{AB})$ and the parameters $t'_i$ according to Eq.~\eqref{eq_eigenvalues}.

Now, if $\det{(T')\le 0}$, then one or three of the $t'_i$ coefficients are less than zero, or some of them are zero. 
In all cases, the maximum eigenvalue of $X_2(\rho_{AB})$ can be written as $\lambda_{\text{max}}=\frac{1}{4}(1+\sum_i |t'_i|)$. 
When $\det{(T') > 0}$, two out of the three coefficients will be negative (or all of them will be positive) making the maximum eigenvalue the one that maximizes the combination $|t'_i|+|t'_j|-|t'_k|$, thus concluding the demonstration.

\vspace{5mm}

\bibliography{bibliography}

\begin{thebibliography}{32}%
\makeatletter
\providecommand \@ifxundefined [1]{%
 \@ifx{#1\undefined}
}%
\providecommand \@ifnum [1]{%
 \ifnum #1\expandafter \@firstoftwo
 \else \expandafter \@secondoftwo
 \fi
}%
\providecommand \@ifx [1]{%
 \ifx #1\expandafter \@firstoftwo
 \else \expandafter \@secondoftwo
 \fi
}%
\providecommand \natexlab [1]{#1}%
\providecommand \enquote  [1]{``#1''}%
\providecommand \bibnamefont  [1]{#1}%
\providecommand \bibfnamefont [1]{#1}%
\providecommand \citenamefont [1]{#1}%
\providecommand \href@noop [0]{\@secondoftwo}%
\providecommand \href [0]{\begingroup \@sanitize@url \@href}%
\providecommand \@href[1]{\@@startlink{#1}\@@href}%
\providecommand \@@href[1]{\endgroup#1\@@endlink}%
\providecommand \@sanitize@url [0]{\catcode `\\12\catcode `\$12\catcode
  `\&12\catcode `\#12\catcode `\^12\catcode `\_12\catcode `\%12\relax}%
\providecommand \@@startlink[1]{}%
\providecommand \@@endlink[0]{}%
\providecommand \url  [0]{\begingroup\@sanitize@url \@url }%
\providecommand \@url [1]{\endgroup\@href {#1}{\urlprefix }}%
\providecommand \urlprefix  [0]{URL }%
\providecommand \Eprint [0]{\href }%
\providecommand \doibase [0]{https://doi.org/}%
\providecommand \selectlanguage [0]{\@gobble}%
\providecommand \bibinfo  [0]{\@secondoftwo}%
\providecommand \bibfield  [0]{\@secondoftwo}%
\providecommand \translation [1]{[#1]}%
\providecommand \BibitemOpen [0]{}%
\providecommand \bibitemStop [0]{}%
\providecommand \bibitemNoStop [0]{.\EOS\space}%
\providecommand \EOS [0]{\spacefactor3000\relax}%
\providecommand \BibitemShut  [1]{\csname bibitem#1\endcsname}%
\let\auto@bib@innerbib\@empty
\bibitem [{\citenamefont {Bourennane}\ \emph {et~al.}(2004)\citenamefont
  {Bourennane}, \citenamefont {Eibl}, \citenamefont {Kurtsiefer}, \citenamefont
  {Gaertner}, \citenamefont {Weinfurter}, \citenamefont {G{\"u}hne},
  \citenamefont {Hyllus}, \citenamefont {Bru{\ss}}, \citenamefont
  {Lewenstein},\ and\ \citenamefont {Sanpera}}]{bourennane2004experimental}%
  \BibitemOpen
  \bibfield  {author} {\bibinfo {author} {\bibfnamefont {M.}~\bibnamefont
  {Bourennane}}, \bibinfo {author} {\bibfnamefont {M.}~\bibnamefont {Eibl}},
  \bibinfo {author} {\bibfnamefont {C.}~\bibnamefont {Kurtsiefer}}, \bibinfo
  {author} {\bibfnamefont {S.}~\bibnamefont {Gaertner}}, \bibinfo {author}
  {\bibfnamefont {H.}~\bibnamefont {Weinfurter}}, \bibinfo {author}
  {\bibfnamefont {O.}~\bibnamefont {G{\"u}hne}}, \bibinfo {author}
  {\bibfnamefont {P.}~\bibnamefont {Hyllus}}, \bibinfo {author} {\bibfnamefont
  {D.}~\bibnamefont {Bru{\ss}}}, \bibinfo {author} {\bibfnamefont
  {M.}~\bibnamefont {Lewenstein}},\ and\ \bibinfo {author} {\bibfnamefont
  {A.}~\bibnamefont {Sanpera}},\ }\bibfield  {title} {\bibinfo {title}
  {Experimental detection of multipartite entanglement using witness
  operators},\ }\href@noop {} {\bibfield  {journal} {\bibinfo  {journal} {Phys.
  Rev. Lett.}\ }\textbf {\bibinfo {volume} {92}},\ \bibinfo {pages} {087902}
  (\bibinfo {year} {2004})}\BibitemShut {NoStop}%
\bibitem [{\citenamefont {Weilenmann}\ \emph {et~al.}(2020)\citenamefont
  {Weilenmann}, \citenamefont {Dive}, \citenamefont {Trillo}, \citenamefont
  {Aguilar},\ and\ \citenamefont {Navascu{\'e}s}}]{weilenmann2020entanglement}%
  \BibitemOpen
  \bibfield  {author} {\bibinfo {author} {\bibfnamefont {M.}~\bibnamefont
  {Weilenmann}}, \bibinfo {author} {\bibfnamefont {B.}~\bibnamefont {Dive}},
  \bibinfo {author} {\bibfnamefont {D.}~\bibnamefont {Trillo}}, \bibinfo
  {author} {\bibfnamefont {E.~A.}\ \bibnamefont {Aguilar}},\ and\ \bibinfo
  {author} {\bibfnamefont {M.}~\bibnamefont {Navascu{\'e}s}},\ }\bibfield
  {title} {\bibinfo {title} {Entanglement detection beyond measuring
  fidelities},\ }\href@noop {} {\bibfield  {journal} {\bibinfo  {journal}
  {Phys. Rev. Lett.}\ }\textbf {\bibinfo {volume} {124}},\ \bibinfo {pages}
  {200502} (\bibinfo {year} {2020})}\BibitemShut {NoStop}%
\bibitem [{\citenamefont {G{\"u}hne}\ \emph {et~al.}(2020)\citenamefont
  {G{\"u}hne}, \citenamefont {Mao},\ and\ \citenamefont
  {Yu}}]{guhne2020geometry}%
  \BibitemOpen
  \bibfield  {author} {\bibinfo {author} {\bibfnamefont {O.}~\bibnamefont
  {G{\"u}hne}}, \bibinfo {author} {\bibfnamefont {Y.}~\bibnamefont {Mao}},\
  and\ \bibinfo {author} {\bibfnamefont {X.-D.}\ \bibnamefont {Yu}},\
  }\bibfield  {title} {\bibinfo {title} {Geometry of faithful entanglement},\
  }\href@noop {} {\bibfield  {journal} {\bibinfo  {journal} {arXiv preprint
  arXiv:2008.05961}\ } (\bibinfo {year} {2020})}\BibitemShut {NoStop}%
\bibitem [{\citenamefont {Zhan}\ and\ \citenamefont
  {Lo}(2020)}]{zhan2020detecting}%
  \BibitemOpen
  \bibfield  {author} {\bibinfo {author} {\bibfnamefont {Y.}~\bibnamefont
  {Zhan}}\ and\ \bibinfo {author} {\bibfnamefont {H.-K.}\ \bibnamefont {Lo}},\
  }\bibfield  {title} {\bibinfo {title} {Detecting entanglement in unfaithful
  states},\ }\href@noop {} {\bibfield  {journal} {\bibinfo  {journal} {arXiv
  preprint arXiv:2010.06054}\ } (\bibinfo {year} {2020})}\BibitemShut {NoStop}%
\bibitem [{\citenamefont {Lang}\ and\ \citenamefont
  {Caves}(2010)}]{lang2010quantum}%
  \BibitemOpen
  \bibfield  {author} {\bibinfo {author} {\bibfnamefont {M.~D.}\ \bibnamefont
  {Lang}}\ and\ \bibinfo {author} {\bibfnamefont {C.~M.}\ \bibnamefont
  {Caves}},\ }\bibfield  {title} {\bibinfo {title} {Quantum discord and the
  geometry of {Bell}-diagonal states},\ }\href@noop {} {\bibfield  {journal}
  {\bibinfo  {journal} {Phys. Rev. Lett.}\ }\textbf {\bibinfo {volume} {105}},\
  \bibinfo {pages} {150501} (\bibinfo {year} {2010})}\BibitemShut {NoStop}%
\bibitem [{\citenamefont {{H. L. Albrecht Q. and D. F.
  Mundarain}}(2020)}]{Albrecht2020maximal}%
  \BibitemOpen
  \bibfield  {author} {\bibinfo {author} {\bibnamefont {{H. L. Albrecht Q. and
  D. F. Mundarain}}},\ }\bibfield  {title} {\bibinfo {title} {Maximal singlet
  fraction vs. maximal achievable fidelity as proper entanglement measures},\
  }\href@noop {} {\bibfield  {journal} {\bibinfo  {journal} {arXiv preprint
  arXiv:2008.13063}\ } (\bibinfo {year} {2020})}\BibitemShut {NoStop}%
\bibitem [{\citenamefont {Gharibian}(2010)}]{gharibian2010strong}%
  \BibitemOpen
  \bibfield  {author} {\bibinfo {author} {\bibfnamefont {S.}~\bibnamefont
  {Gharibian}},\ }\bibfield  {title} {\bibinfo {title} {Strong {NP}-hardness of
  the quantum separability problem},\ }\href@noop {} {\bibfield  {journal}
  {\bibinfo  {journal} {Quantum Inf. Comput.}\ }\textbf {\bibinfo {volume}
  {10}},\ \bibinfo {pages} {343} (\bibinfo {year} {2010})}\BibitemShut
  {NoStop}%
\bibitem [{\citenamefont {G{\"u}hne}\ and\ \citenamefont
  {T{\'o}th}(2009)}]{guhne2009entanglement}%
  \BibitemOpen
  \bibfield  {author} {\bibinfo {author} {\bibfnamefont {O.}~\bibnamefont
  {G{\"u}hne}}\ and\ \bibinfo {author} {\bibfnamefont {G.}~\bibnamefont
  {T{\'o}th}},\ }\bibfield  {title} {\bibinfo {title} {Entanglement
  detection},\ }\href@noop {} {\bibfield  {journal} {\bibinfo  {journal} {Phys.
  Rep.}\ }\textbf {\bibinfo {volume} {474}},\ \bibinfo {pages} {1} (\bibinfo
  {year} {2009})}\BibitemShut {NoStop}%
\bibitem [{\citenamefont {Terhal}(2002)}]{terhal2002detecting}%
  \BibitemOpen
  \bibfield  {author} {\bibinfo {author} {\bibfnamefont {B.~M.}\ \bibnamefont
  {Terhal}},\ }\bibfield  {title} {\bibinfo {title} {Detecting quantum
  entanglement},\ }\href@noop {} {\bibfield  {journal} {\bibinfo  {journal}
  {Theor. Comput. Sci.}\ }\textbf {\bibinfo {volume} {287}},\ \bibinfo {pages}
  {313} (\bibinfo {year} {2002})}\BibitemShut {NoStop}%
\bibitem [{\citenamefont {Horodecki}\ \emph
  {et~al.}(1996{\natexlab{a}})\citenamefont {Horodecki}, \citenamefont
  {Horodecki},\ and\ \citenamefont {Horodecki}}]{HORODECKI19961}%
  \BibitemOpen
  \bibfield  {author} {\bibinfo {author} {\bibfnamefont {M.}~\bibnamefont
  {Horodecki}}, \bibinfo {author} {\bibfnamefont {P.}~\bibnamefont
  {Horodecki}},\ and\ \bibinfo {author} {\bibfnamefont {R.}~\bibnamefont
  {Horodecki}},\ }\bibfield  {title} {\bibinfo {title} {Separability of mixed
  states: necessary and sufficient conditions},\ }\href
  {https://doi.org/https://doi.org/10.1016/S0375-9601(96)00706-2} {\bibfield
  {journal} {\bibinfo  {journal} {Phys. Lett. A}\ }\textbf {\bibinfo {volume}
  {223}},\ \bibinfo {pages} {1 } (\bibinfo {year}
  {1996}{\natexlab{a}})}\BibitemShut {NoStop}%
\bibitem [{\citenamefont {Wootters}(2001)}]{wootters2001entanglement}%
  \BibitemOpen
  \bibfield  {author} {\bibinfo {author} {\bibfnamefont {W.~K.}\ \bibnamefont
  {Wootters}},\ }\bibfield  {title} {\bibinfo {title} {Entanglement of
  formation and concurrence.},\ }\href@noop {} {\bibfield  {journal} {\bibinfo
  {journal} {Quantum Inf. Comput.}\ }\textbf {\bibinfo {volume} {1}},\ \bibinfo
  {pages} {27} (\bibinfo {year} {2001})}\BibitemShut {NoStop}%
\bibitem [{\citenamefont {Horodecki}\ \emph
  {et~al.}(1996{\natexlab{b}})\citenamefont {Horodecki}, \citenamefont
  {Horodecki},\ and\ \citenamefont {Horodecki}}]{horodecki1996teleportation}%
  \BibitemOpen
  \bibfield  {author} {\bibinfo {author} {\bibfnamefont {R.}~\bibnamefont
  {Horodecki}}, \bibinfo {author} {\bibfnamefont {M.}~\bibnamefont
  {Horodecki}},\ and\ \bibinfo {author} {\bibfnamefont {P.}~\bibnamefont
  {Horodecki}},\ }\bibfield  {title} {\bibinfo {title} {Teleportation, {Bell}'s
  inequalities and inseparability},\ }\href@noop {} {\bibfield  {journal}
  {\bibinfo  {journal} {Phys. Lett. A}\ }\textbf {\bibinfo {volume} {222}},\
  \bibinfo {pages} {21} (\bibinfo {year} {1996}{\natexlab{b}})}\BibitemShut
  {NoStop}%
\bibitem [{\citenamefont {Horodecki}\ \emph {et~al.}(1999)\citenamefont
  {Horodecki}, \citenamefont {Horodecki},\ and\ \citenamefont
  {Horodecki}}]{horodecki1999general}%
  \BibitemOpen
  \bibfield  {author} {\bibinfo {author} {\bibfnamefont {M.}~\bibnamefont
  {Horodecki}}, \bibinfo {author} {\bibfnamefont {P.}~\bibnamefont
  {Horodecki}},\ and\ \bibinfo {author} {\bibfnamefont {R.}~\bibnamefont
  {Horodecki}},\ }\bibfield  {title} {\bibinfo {title} {General teleportation
  channel, singlet fraction, and quasidistillation},\ }\href@noop {} {\bibfield
   {journal} {\bibinfo  {journal} {Phys. Rev. A}\ }\textbf {\bibinfo {volume}
  {60}},\ \bibinfo {pages} {1888} (\bibinfo {year} {1999})}\BibitemShut
  {NoStop}%
\bibitem [{\citenamefont {Grondalski}\ \emph {et~al.}(2002)\citenamefont
  {Grondalski}, \citenamefont {Etlinger},\ and\ \citenamefont
  {James}}]{grondalski2002fully}%
  \BibitemOpen
  \bibfield  {author} {\bibinfo {author} {\bibfnamefont {J.}~\bibnamefont
  {Grondalski}}, \bibinfo {author} {\bibfnamefont {D.}~\bibnamefont
  {Etlinger}},\ and\ \bibinfo {author} {\bibfnamefont {D.}~\bibnamefont
  {James}},\ }\bibfield  {title} {\bibinfo {title} {The fully entangled
  fraction as an inclusive measure of entanglement applications},\ }\href@noop
  {} {\bibfield  {journal} {\bibinfo  {journal} {Phys. Lett. A}\ }\textbf
  {\bibinfo {volume} {300}},\ \bibinfo {pages} {573} (\bibinfo {year}
  {2002})}\BibitemShut {NoStop}%
\bibitem [{\citenamefont {Verstraete}\ and\ \citenamefont
  {Verschelde}(2002)}]{verstraete2002fidelity}%
  \BibitemOpen
  \bibfield  {author} {\bibinfo {author} {\bibfnamefont {F.}~\bibnamefont
  {Verstraete}}\ and\ \bibinfo {author} {\bibfnamefont {H.}~\bibnamefont
  {Verschelde}},\ }\bibfield  {title} {\bibinfo {title} {Fidelity of mixed
  states of two qubits},\ }\href@noop {} {\bibfield  {journal} {\bibinfo
  {journal} {Phys. Rev. A}\ }\textbf {\bibinfo {volume} {66}},\ \bibinfo
  {pages} {022307} (\bibinfo {year} {2002})}\BibitemShut {NoStop}%
\bibitem [{\citenamefont {Bennett}\ \emph {et~al.}(1996)\citenamefont
  {Bennett}, \citenamefont {DiVincenzo}, \citenamefont {Smolin},\ and\
  \citenamefont {Wootters}}]{bennett1996mixed}%
  \BibitemOpen
  \bibfield  {author} {\bibinfo {author} {\bibfnamefont {C.~H.}\ \bibnamefont
  {Bennett}}, \bibinfo {author} {\bibfnamefont {D.~P.}\ \bibnamefont
  {DiVincenzo}}, \bibinfo {author} {\bibfnamefont {J.~A.}\ \bibnamefont
  {Smolin}},\ and\ \bibinfo {author} {\bibfnamefont {W.~K.}\ \bibnamefont
  {Wootters}},\ }\bibfield  {title} {\bibinfo {title} {Mixed-state entanglement
  and quantum error correction},\ }\href@noop {} {\bibfield  {journal}
  {\bibinfo  {journal} {Phys. Rev. A}\ }\textbf {\bibinfo {volume} {54}},\
  \bibinfo {pages} {3824} (\bibinfo {year} {1996})}\BibitemShut {NoStop}%
\bibitem [{\citenamefont {Badziag}\ \emph {et~al.}(2000)\citenamefont
  {Badziag}, \citenamefont {Horodecki}, \citenamefont {Horodecki},\ and\
  \citenamefont {Horodecki}}]{badziag2000local}%
  \BibitemOpen
  \bibfield  {author} {\bibinfo {author} {\bibfnamefont {P.}~\bibnamefont
  {Badziag}}, \bibinfo {author} {\bibfnamefont {M.}~\bibnamefont {Horodecki}},
  \bibinfo {author} {\bibfnamefont {P.}~\bibnamefont {Horodecki}},\ and\
  \bibinfo {author} {\bibfnamefont {R.}~\bibnamefont {Horodecki}},\ }\bibfield
  {title} {\bibinfo {title} {Local environment can enhance fidelity of quantum
  teleportation},\ }\href@noop {} {\bibfield  {journal} {\bibinfo  {journal}
  {Phys. Rev. A}\ }\textbf {\bibinfo {volume} {62}},\ \bibinfo {pages} {012311}
  (\bibinfo {year} {2000})}\BibitemShut {NoStop}%
\bibitem [{\citenamefont {Yu}\ \emph {et~al.}(2020)\citenamefont {Yu},
  \citenamefont {Simnacher}, \citenamefont {Nguyen},\ and\ \citenamefont
  {G{\"u}hne}}]{yu2020quantuminspired}%
  \BibitemOpen
  \bibfield  {author} {\bibinfo {author} {\bibfnamefont {X.-D.}\ \bibnamefont
  {Yu}}, \bibinfo {author} {\bibfnamefont {T.}~\bibnamefont {Simnacher}},
  \bibinfo {author} {\bibfnamefont {H.~C.}\ \bibnamefont {Nguyen}},\ and\
  \bibinfo {author} {\bibfnamefont {O.}~\bibnamefont {G{\"u}hne}},\ }\bibfield
  {title} {\bibinfo {title} {Quantum-inspired hierarchy for rank-constrained
  optimization},\ }\href@noop {} {\bibfield  {journal} {\bibinfo  {journal}
  {arXiv preprint arXiv:2012.00554}\ } (\bibinfo {year} {2020})}\BibitemShut
  {NoStop}%
\bibitem [{\citenamefont {Antonelli}\ \emph {et~al.}(2011)\citenamefont
  {Antonelli}, \citenamefont {Shtaif},\ and\ \citenamefont
  {Brodsky}}]{antonelli2011sudden}%
  \BibitemOpen
  \bibfield  {author} {\bibinfo {author} {\bibfnamefont {C.}~\bibnamefont
  {Antonelli}}, \bibinfo {author} {\bibfnamefont {M.}~\bibnamefont {Shtaif}},\
  and\ \bibinfo {author} {\bibfnamefont {M.}~\bibnamefont {Brodsky}},\
  }\bibfield  {title} {\bibinfo {title} {Sudden death of entanglement induced
  by polarization mode dispersion},\ }\href@noop {} {\bibfield  {journal}
  {\bibinfo  {journal} {Phys. Rev. Lett.}\ }\textbf {\bibinfo {volume} {106}},\
  \bibinfo {pages} {080404} (\bibinfo {year} {2011})}\BibitemShut {NoStop}%
\bibitem [{\citenamefont {Kirby}\ \emph {et~al.}(2019)\citenamefont {Kirby},
  \citenamefont {Jones},\ and\ \citenamefont {Brodsky}}]{kirby2019effect}%
  \BibitemOpen
  \bibfield  {author} {\bibinfo {author} {\bibfnamefont {B.~T.}\ \bibnamefont
  {Kirby}}, \bibinfo {author} {\bibfnamefont {D.~E.}\ \bibnamefont {Jones}},\
  and\ \bibinfo {author} {\bibfnamefont {M.}~\bibnamefont {Brodsky}},\
  }\bibfield  {title} {\bibinfo {title} {Effect of polarization dependent loss
  on the quality of transmitted polarization entanglement},\ }\href@noop {}
  {\bibfield  {journal} {\bibinfo  {journal} {J. Light. Technol.}\ }\textbf
  {\bibinfo {volume} {37}},\ \bibinfo {pages} {95} (\bibinfo {year}
  {2019})}\BibitemShut {NoStop}%
\bibitem [{\citenamefont {Verstraete}\ \emph {et~al.}(2001)\citenamefont
  {Verstraete}, \citenamefont {Dehaene},\ and\ \citenamefont
  {DeMoor}}]{verstraete2001local}%
  \BibitemOpen
  \bibfield  {author} {\bibinfo {author} {\bibfnamefont {F.}~\bibnamefont
  {Verstraete}}, \bibinfo {author} {\bibfnamefont {J.}~\bibnamefont
  {Dehaene}},\ and\ \bibinfo {author} {\bibfnamefont {B.}~\bibnamefont
  {DeMoor}},\ }\bibfield  {title} {\bibinfo {title} {Local filtering operations
  on two qubits},\ }\href@noop {} {\bibfield  {journal} {\bibinfo  {journal}
  {Phys. Rev. A}\ }\textbf {\bibinfo {volume} {64}},\ \bibinfo {pages} {010101}
  (\bibinfo {year} {2001})}\BibitemShut {NoStop}%
\bibitem [{\citenamefont {Fiorentino}\ \emph {et~al.}(2002)\citenamefont
  {Fiorentino}, \citenamefont {Voss}, \citenamefont {Sharping},\ and\
  \citenamefont {Kumar}}]{fiorentino2002all}%
  \BibitemOpen
  \bibfield  {author} {\bibinfo {author} {\bibfnamefont {M.}~\bibnamefont
  {Fiorentino}}, \bibinfo {author} {\bibfnamefont {P.~L.}\ \bibnamefont
  {Voss}}, \bibinfo {author} {\bibfnamefont {J.~E.}\ \bibnamefont {Sharping}},\
  and\ \bibinfo {author} {\bibfnamefont {P.}~\bibnamefont {Kumar}},\ }\bibfield
   {title} {\bibinfo {title} {All-fiber photon-pair source for quantum
  communications},\ }\href@noop {} {\bibfield  {journal} {\bibinfo  {journal}
  {IEEE Photonics Technol. Lett.}\ }\textbf {\bibinfo {volume} {14}},\ \bibinfo
  {pages} {983} (\bibinfo {year} {2002})}\BibitemShut {NoStop}%
\bibitem [{\citenamefont {Wang}\ and\ \citenamefont
  {Kanter}(2009)}]{wang2009robust}%
  \BibitemOpen
  \bibfield  {author} {\bibinfo {author} {\bibfnamefont {S.~X.}\ \bibnamefont
  {Wang}}\ and\ \bibinfo {author} {\bibfnamefont {G.~S.}\ \bibnamefont
  {Kanter}},\ }\bibfield  {title} {\bibinfo {title} {Robust multiwavelength
  all-fiber source of polarization-entangled photons with built-in analyzer
  alignment signal},\ }\href@noop {} {\bibfield  {journal} {\bibinfo  {journal}
  {IEEE J. Sel. Top. Quantum Electron.}\ }\textbf {\bibinfo {volume} {15}},\
  \bibinfo {pages} {1733} (\bibinfo {year} {2009})}\BibitemShut {NoStop}%
\bibitem [{Note1()}]{Note1}%
  \BibitemOpen
  \bibinfo {note} {Full state tomography of two qubits is achievable from the
  statistics of only $9$ measurement settings if two detectors are used per
  photon \cite {altepeter2005photonic}. However, for convenience we use a
  single detector per photon and take $36$ total measurements, corresponding to
  all pairwise combinations of both eigenstates of each Pauli
  operator.}\BibitemShut {Stop}%
\bibitem [{\citenamefont {Altepeter}\ \emph {et~al.}(2005)\citenamefont
  {Altepeter}, \citenamefont {Jeffrey},\ and\ \citenamefont
  {Kwiat}}]{altepeter2005photonic}%
  \BibitemOpen
  \bibfield  {author} {\bibinfo {author} {\bibfnamefont {J.~B.}\ \bibnamefont
  {Altepeter}}, \bibinfo {author} {\bibfnamefont {E.~R.}\ \bibnamefont
  {Jeffrey}},\ and\ \bibinfo {author} {\bibfnamefont {P.~G.}\ \bibnamefont
  {Kwiat}},\ }\bibfield  {title} {\bibinfo {title} {Photonic state
  tomography},\ }\href@noop {} {\bibfield  {journal} {\bibinfo  {journal} {Adv.
  Atom. Mol. Opt. Phys.}\ }\textbf {\bibinfo {volume} {52}},\ \bibinfo {pages}
  {105} (\bibinfo {year} {2005})}\BibitemShut {NoStop}%
\bibitem [{\citenamefont {Brodsky}\ \emph {et~al.}(2011)\citenamefont
  {Brodsky}, \citenamefont {George}, \citenamefont {Antonelli},\ and\
  \citenamefont {Shtaif}}]{brodsky2011loss}%
  \BibitemOpen
  \bibfield  {author} {\bibinfo {author} {\bibfnamefont {M.}~\bibnamefont
  {Brodsky}}, \bibinfo {author} {\bibfnamefont {E.~C.}\ \bibnamefont {George}},
  \bibinfo {author} {\bibfnamefont {C.}~\bibnamefont {Antonelli}},\ and\
  \bibinfo {author} {\bibfnamefont {M.}~\bibnamefont {Shtaif}},\ }\bibfield
  {title} {\bibinfo {title} {Loss of polarization entanglement in a fiber-optic
  system with polarization mode dispersion in one optical path},\ }\href@noop
  {} {\bibfield  {journal} {\bibinfo  {journal} {Opt. Lett.}\ }\textbf
  {\bibinfo {volume} {36}},\ \bibinfo {pages} {43} (\bibinfo {year}
  {2011})}\BibitemShut {NoStop}%
\bibitem [{\citenamefont {Shtaif}\ \emph {et~al.}(2011)\citenamefont {Shtaif},
  \citenamefont {Antonelli},\ and\ \citenamefont
  {Brodsky}}]{shtaif2011nonlocal}%
  \BibitemOpen
  \bibfield  {author} {\bibinfo {author} {\bibfnamefont {M.}~\bibnamefont
  {Shtaif}}, \bibinfo {author} {\bibfnamefont {C.}~\bibnamefont {Antonelli}},\
  and\ \bibinfo {author} {\bibfnamefont {M.}~\bibnamefont {Brodsky}},\
  }\bibfield  {title} {\bibinfo {title} {Nonlocal compensation of polarization
  mode dispersion in the transmission of polarization entangled photons},\
  }\href@noop {} {\bibfield  {journal} {\bibinfo  {journal} {Opt. Express}\
  }\textbf {\bibinfo {volume} {19}},\ \bibinfo {pages} {1728} (\bibinfo {year}
  {2011})}\BibitemShut {NoStop}%
\bibitem [{\citenamefont {Jones}\ \emph {et~al.}(2018)\citenamefont {Jones},
  \citenamefont {Kirby},\ and\ \citenamefont {Brodsky}}]{jones2018tuning}%
  \BibitemOpen
  \bibfield  {author} {\bibinfo {author} {\bibfnamefont {D.~E.}\ \bibnamefont
  {Jones}}, \bibinfo {author} {\bibfnamefont {B.~T.}\ \bibnamefont {Kirby}},\
  and\ \bibinfo {author} {\bibfnamefont {M.}~\bibnamefont {Brodsky}},\
  }\bibfield  {title} {\bibinfo {title} {Tuning quantum channels to maximize
  polarization entanglement for telecom photon pairs},\ }\href@noop {}
  {\bibfield  {journal} {\bibinfo  {journal} {npj Quantum Inf.}\ }\textbf
  {\bibinfo {volume} {4}},\ \bibinfo {pages} {58} (\bibinfo {year}
  {2018})}\BibitemShut {NoStop}%
\bibitem [{\citenamefont {Jones}\ \emph
  {et~al.}(2020{\natexlab{a}})\citenamefont {Jones}, \citenamefont {Kirby},
  \citenamefont {Riccardi}, \citenamefont {Antonelli},\ and\ \citenamefont
  {Brodsky}}]{jones2020exploring}%
  \BibitemOpen
  \bibfield  {author} {\bibinfo {author} {\bibfnamefont {D.~E.}\ \bibnamefont
  {Jones}}, \bibinfo {author} {\bibfnamefont {B.~T.}\ \bibnamefont {Kirby}},
  \bibinfo {author} {\bibfnamefont {G.}~\bibnamefont {Riccardi}}, \bibinfo
  {author} {\bibfnamefont {C.}~\bibnamefont {Antonelli}},\ and\ \bibinfo
  {author} {\bibfnamefont {M.}~\bibnamefont {Brodsky}},\ }\bibfield  {title}
  {\bibinfo {title} {Exploring classical correlations in noise to recover
  quantum information using local filtering},\ }\href@noop {} {\bibfield
  {journal} {\bibinfo  {journal} {New J. Phys.}\ }\textbf {\bibinfo {volume}
  {22}},\ \bibinfo {pages} {073037} (\bibinfo {year}
  {2020}{\natexlab{a}})}\BibitemShut {NoStop}%
\bibitem [{\citenamefont {Riccardi}\ \emph {et~al.}(2021)\citenamefont
  {Riccardi}, \citenamefont {Antonelli}, \citenamefont {Jones},\ and\
  \citenamefont {Brodsky}}]{riccardi2021simultaneous}%
  \BibitemOpen
  \bibfield  {author} {\bibinfo {author} {\bibfnamefont {G.}~\bibnamefont
  {Riccardi}}, \bibinfo {author} {\bibfnamefont {C.}~\bibnamefont {Antonelli}},
  \bibinfo {author} {\bibfnamefont {D.~E.}\ \bibnamefont {Jones}},\ and\
  \bibinfo {author} {\bibfnamefont {M.}~\bibnamefont {Brodsky}},\ }\bibfield
  {title} {\bibinfo {title} {Simultaneous decoherence and mode filtering in
  quantum channels: Theory and experiment},\ }\href@noop {} {\bibfield
  {journal} {\bibinfo  {journal} {Phys. Rev. Applied}\ }\textbf {\bibinfo
  {volume} {15}},\ \bibinfo {pages} {014060} (\bibinfo {year}
  {2021})}\BibitemShut {NoStop}%
\bibitem [{\citenamefont {Jones}\ \emph
  {et~al.}(2020{\natexlab{b}})\citenamefont {Jones}, \citenamefont {Kirby},
  \citenamefont {Riccardi}, \citenamefont {Antonelli},\ and\ \citenamefont
  {Brodsky}}]{jones2020experimental}%
  \BibitemOpen
  \bibfield  {author} {\bibinfo {author} {\bibfnamefont {D.~E.}\ \bibnamefont
  {Jones}}, \bibinfo {author} {\bibfnamefont {B.~T.}\ \bibnamefont {Kirby}},
  \bibinfo {author} {\bibfnamefont {G.}~\bibnamefont {Riccardi}}, \bibinfo
  {author} {\bibfnamefont {C.}~\bibnamefont {Antonelli}},\ and\ \bibinfo
  {author} {\bibfnamefont {M.}~\bibnamefont {Brodsky}},\ }\href@noop {} {\emph
  {\bibinfo {title} {Experimental Procedure to Characterize the Effects of
  Filtering and Decoherence on Polarization Entangled Photons}}},\ \bibinfo
  {type} {Tech. Rep.}\ (\bibinfo  {institution} {DEVCOM U.S. Army Research
  Laboratory Adelphi United States},\ \bibinfo {year} {2020})\BibitemShut
  {NoStop}%
\bibitem [{Note2()}]{Note2}%
  \BibitemOpen
  \bibinfo {note} {The matrices in Figs. \ref {expfig1}-\ref {expfig2} are
  expressed in the basis of the filter $\protect \vec n_{A}$, and the $x$-axis
  in Stokes space is defined as $\vert H \rangle $. This basis allows for an
  intuitive understanding of how the filters $\protect \vec n_{A}$ and
  $\protect \vec n_{B}$ alter the density matrix; however, we note that the HV
  basis is rotated by $\pi /2$ (for both qubits) relative to the basis of the
  density matrix described by Eq. (16).}\BibitemShut {Stop}%
\end{thebibliography}%

\end{document}